# Regularization by ε-metric. II. Limit ε = + 0

V.D. Ivashchuk

VNIIMS & RUDN University,

Moscow, Russia

*Abstract.*

In a wide class of propagators regularized by the ε-metric [1], the R-operation is formulated. It is proved that the limit of renormalized Feynman integrals exists and is covariant. Possible applications in gravity are discussed.[*)]

1. **Introduction.** In the previous work of the author [1], the regularization of propagators was introduced using a complex ε-metric in $\boldsymbol{R}^D$ ($D \geq 2$):

$$\left(\eta^{\varepsilon}_{\alpha\beta}\right) = \begin{pmatrix} e^{-i\varepsilon} & 0 & \dots & \dots & 0 \\ 0 & -1 & \dots & \dots & 0 \\ \dots & \dots & \ddots & \dots & \vdots \\ \dots & \dots & \dots & -1 & 0 \\ 0 & \dots & \dots & 0 & -1 \end{pmatrix}, \; 0 < \varepsilon < 2\pi. \qquad (1)$$

ε = +0 corresponds to the Minkowski space[time] with the correct direction of the time arrow, ε = π corresponds to the Euclidean space.

In this case, the propagator (ε-correlator) of a free massive scalar field regularized with (1) in the momentum representation has the form [1]:

$$(\underline{\varphi\varphi})_{\varepsilon}(p) \; = \frac{i \, e^{i\varepsilon/2}}{p_{\varepsilon}^2 - m^2}, \qquad (2)$$

_______________________________________________________________________________

where $p_\varepsilon^2 = e^{i\varepsilon} p_0^2 - \boldsymbol{p}^2$, and (2) admits the α-representation [1]:

$$\frac{i e^{i\varepsilon/2}}{p_\varepsilon^2 - m^2} = \int_0^\infty d\alpha \, \exp\left(i \alpha \, e^{-i\varepsilon/2} \, (p_\varepsilon^2 - m^2)\right),$$

where $m^2 > 0$, $\varepsilon \in (0, 2\pi)$, $p \in \boldsymbol{R}^D$. (As in [1], all integrals are Lebesgue integrals [2]). In [1], a wide class of proper ε-correlators was introduced. For convenience of presentation, the definition of the proper correlator is given in section 2.

In section 3 of this paper, we consider an arbitrary Feynman integral made up of proper ε-correlators and corresponding to a connected graph $\Gamma$. If the divergence indices of subgraphs $\Gamma$ are negative, then for $0 < \varepsilon < 2\pi$ and all external momenta, the integral exists and is reduced to a convergent integral over α-variables, the explicit form of which is given (Proposition 1).

The R-operation is formulated in the α-representation. If we take a set of nests [3] composed of arbitrary sets of lines as a renormalization array, then the renormalized integral for $0 < \varepsilon < 2\pi$ and all external momenta exists, generates a regular generalized function that has a covariant limit (in the space of generalized functions) for $\varepsilon \to +0$. Herewith the question of reducing the renormalization array is considered. It turns out that if the ultraviolet [UV] dimensions of the correlators [1] satisfy the inequality

$$-D \leq d_1 < 0,$$

then the renormalized integral does not change if we make the following renormalization array replacement: nests from arbitrary sets of lines $(N(L))$ − forests from strongly connected sets of lines $(F(L_1))$ − forests from strongly connected complete [3] sets of lines $(F(L_{1c}))$.

In section 4, possible applications of regularization by the ε-metric in gravity are discussed.

2. **Proper ε-correlators.** We present the definition of proper ε-correlator and a number of relations from [1].

**Definition 1.** *ε-correlator $(\Phi\Phi)_\varepsilon (p)$, $\varepsilon \in (0, 2\pi)$ is called proper one, if for all $\varepsilon \in (0, 2\pi)$ (and $p \in \boldsymbol{R}^D$) it can be presented as*

$$(\Phi\Phi)_\varepsilon(p) = P(p,\varepsilon) \int_0^\infty d\alpha\, f(i\alpha\, e^{-i\varepsilon/2})\exp\left(i\alpha\, e^{-i\varepsilon/2}(p_\varepsilon^2 - m^2)\right), \qquad (3)$$

$1^0$. a) $f(\alpha)$ is holomorphic in the region $\{\mathrm{Re}\,\alpha > 0\}$ and continuous in $\{\mathrm{Re}\,\alpha \geq 0\}\setminus\{0\}$;

b) $f(\alpha) = O(\alpha^T)$ for $\alpha \to \infty$, $\mathrm{Re}\,\alpha \geq 0$;

c) for some $s > -1$ and all $\delta > 0$ the relations are valid

$$f(\alpha) = O(\alpha^{s-\delta}),$$

and

$$\alpha^{s+\delta} = O(f(\alpha)),$$

when $\alpha \to 0$, $\mathrm{Re}\,\alpha \geq 0$;

$2^0$. $m^2 > 0$;

$3^0$. there exists a holomorphic function $F(z)$ defined in $\mathbb{C}\setminus(-\infty, 0]$, which for $\mathrm{Re}\,z > 0$ is equal to

$$F(z) = \int_0^\infty d\alpha\, f(\alpha) e^{-\alpha z};$$

$4^0$. $P(p,\varepsilon)$ is a polynomial in $p_\alpha$ with coefficients (possibly matrix-valued) smoothly depending on $\varepsilon$, and $P(p,0)$ is covariant.

The integral in (3) converges due to the conditions $1^\circ$, $2^\circ$ and the relation:

$$\left|\exp\left(ie^{-i\varepsilon/2}\alpha(p_\varepsilon^2 - m^2)\right)\right| = \exp(-\alpha \sin(\varepsilon/2)(p_E^2 + m^2)). \qquad (4)$$

As was established in [1], the quantity

$$d = \dim P - 2 - 2s, \qquad (5)$$

where dim P is the highest degree of the polynomial in (3), characterizes the behavior of (3) at large momenta and is called the ultraviolet dimension of the $\varepsilon$-correlator (3).

We also present the majorizing inequalities from [1]:

$$\left(\sin\frac{\varepsilon}{2}\right)(k_E^2 + m^2) \leq |k_\varepsilon^2 - m^2| \leq k_E^2 + m^2, \tag{6}$$

where $k_E^2 = k_0^2 + \bar{k}^2$ is the Euclidean square [of the vector].

3. **Feynman Integrals. R-operation. The limit $\varepsilon = +0$.** Consider the Feynman integral represented by a connected graph $\Gamma$. Let each line of the graph presents a proper $\varepsilon$-correlator (3). In the simplest case of the $\varepsilon$-correlator (2), the Feynman integral exists for $\varepsilon \in (0, 2\pi)$ if it exists in the Euclidean case ($\varepsilon = \pi$). This follows from the left inequality (6). We formulate a similar statement for proper $\varepsilon$-correlators (3) in a weakened version.

**Proposition 1.** *Given a Feynman integral composed of proper $\varepsilon$-correlators (3) and represented by a connected graph $\Gamma$. Let all divergence indices of subgraphs $\Gamma$ be negative [($\omega_\gamma < 0$)]. Then the Feynman integral exists for $0 < \varepsilon < 2\pi$ and all external momenta q and is represented as:*

$$I(q, \varepsilon) = (4\pi)^{-DP/2} \exp(i(\varepsilon - \pi)(D-2)/4) \times$$

$$\times \int_0^\infty d\alpha \left(\prod_{l=1}^{|L|} f_l\left(\alpha_l i e^{-i\varepsilon/2}\right)\right) \exp\left(-\sum_{l=1}^{|L|} m_l^2 \alpha_l i e^{-i\varepsilon/2}\right) \prod_{l=1}^{|L|} P_l\left(\frac{\partial}{2i\partial\xi_l}, \varepsilon\right) f(\alpha, q, \xi, \varepsilon)_{|\xi=0}, \tag{7}$$

*where*

$$f(\alpha, q, \xi, \varepsilon) = (D(\alpha))^{-D/2} \exp\{i(e^{-i\varepsilon/2} A(q, \alpha, \varepsilon) - 2B(q, \xi, \alpha) - K(\xi, \alpha, \varepsilon) e^{i\varepsilon/2})(D(\alpha))^{-1}\}, \tag{8}$$

$B(q, \xi, \alpha), D(\alpha)$ *are defined in [3] (p. 76 – 77), and $A(q, \alpha, \varepsilon)$ and $K(\xi, \alpha, \varepsilon)$ differ from those listed in [3] (p. 76 – 78) by replacing the Minkowski metric with the $\varepsilon$-metric. In (7) $|L|$ is the number of lines in $\Gamma$, P is the number of loops in $\Gamma$.*

We emphasize that in (8) $q_l$ are momenta, and $\xi_l$ are "coordinates", so that by virtue of the regularization prescription [1]:

$$(q_l)_\varepsilon^2 = e^{i\varepsilon}(q_{l0})^2 - \bar{q}_l^2;$$

$$(\xi_l)_\varepsilon^2 = e^{-i\varepsilon}(\xi_l^0)^2 - \bar{\xi}_l^2;$$

$$q_l \xi_m = q_{l\alpha} \xi_m^\alpha.$$

**Proof.** The Feynman integral with proper ε-correlators (3) has the form:
$$I(q, \varepsilon) = \int dk\, F(k, q, \varepsilon) = \int dk \int d\alpha\, \Phi(\alpha, k, q, \varepsilon), \tag{9}$$

where $q$ is the set of external momenta; $\varepsilon \in (0, 2\pi)$. The existence of the integral (9) for $\varepsilon \in (0, 2\pi)$, as well as the mixed integral

$$\int dk d\alpha\, \Phi(k, q, \varepsilon)$$

and the repeated one

$$\int d\alpha \int dk\, \Phi(\alpha, k, q, \varepsilon)$$

and their equality to each other follows, for example, from the existence of the repeated integral

$$\int d\alpha \int dk |\Phi(\alpha, k, q, \varepsilon)| \tag{10}$$

(see [2], p. 318). The existence of [integral] (10) follows from the behavior $f_l(\alpha_l)$ (see Definition 1) and relation (4). Indeed, the momentum integral in (10) exists and has the Euclidean form (in $D(\alpha), K(\xi, \alpha, \varepsilon), A(q, \alpha, \varepsilon), B(q, \xi, \alpha)$, $\alpha$ is replaced by $\alpha \sin\frac{\varepsilon}{2}$ ). In this case, similar to the analysis carried out in [3] (p. 134), it is established that the singular point $\alpha = 0$ is integrable (since $\omega_\gamma < 0$).

At high α the convergence is provided by the cutting factor $\exp\left(-\sum_{l=1}^{|L|} m_l^2 \alpha_l \sin\frac{\varepsilon}{2}\right)$. As for the calculations that lead to (7) - (8), they are easily carried out using combinatorial relations from [3] (p. 78-90) and the formula

$$\int \frac{d^D k}{(2\pi)^D} \exp\left(ie^{-i\varepsilon/2} \cdot k_\varepsilon^2\right) = \frac{\exp(i(\varepsilon - \pi)(D-2)/4)}{(4\pi)^{D/2}},$$

where $0 < \varepsilon < 2\pi$.

In general, the integral (7) may diverge. We formulate an $R$-operation that matches the integral (7) with a convergent integral for $0 < \varepsilon < 2\pi$. It is obtained from (7) by replacing

$$f \to Rf,$$

where

$$(Rf)(\alpha, q, \xi, \varepsilon) = (Rf)_{\mathcal{A}(X)}(\alpha, q, \xi, \varepsilon) =$$

$$\left\{ \prod_{\gamma \in \mathcal{L}} I_\gamma + \sum_{A \in \mathcal{A}(X)} \prod_{\gamma \in A} (-M_\gamma) \prod_{\gamma \notin A} I_\gamma \right\} \times$$

$$\times \prod_{\gamma \in \mathcal{L}} \sigma_\gamma^{-1-[\sum_{l \in \gamma} d_l]} f(\beta(\alpha, \sigma), q, \zeta(\alpha, \xi), \varepsilon). \tag{11}$$

Here $\mathcal{L} = P(L) \setminus \{\emptyset\}$ is the set of non-empty sets of lines of graph $\Gamma$; $X \in \mathcal{L}$ and $\mathcal{A}(X)$ is the set of non-empty forests ($F(X)$) or nests ($N(X)$), made up of $X$ [3].

As $X$ is usually used $\mathcal{L}, \mathcal{L}_1, \mathcal{L}_{1c}, \mathcal{L}_1^+, \mathcal{L}_{1c}^+$, where $\mathcal{L}_1$ is the subset of $\mathcal{L}$, consisting of all strongly connected sets of lines, $\mathcal{L}_{1c}$ is the subset (of $\mathcal{L}$) consisting of all strongly connected complete sets of lines. $\mathcal{L}_1^+ = \{\gamma | \gamma \in \mathcal{L}_1, \omega(\gamma) \geq 0\}$ and analogous definition for $\mathcal{L}_{1c}^+$. Operators $I_\gamma, M_\gamma$ are defined in [3] (p. 110) and

$$\beta_l(\alpha, \sigma) = \alpha_l \prod_{\gamma \ni l} \sigma_\gamma^2 ;$$

and

$$\zeta_l(\xi, \sigma) = \xi_l \prod_{\gamma \ni l} \sigma_\gamma.$$

In (11) $\left[\sum_{l \in \gamma} d_l\right]$ is integer part of the sum of dimensions of $\varepsilon$-correlators. We call $\mathcal{A}(X)$ as a renormalization array. The following proposition is valid

**Proposition 2.** If $d_l < 0, \forall l \in L,$ (12)

then $\quad (Rf)_{N(\mathcal{L})} = (Rf)_{F(\mathcal{L}_1)} ;$ (12a)

if $\quad -D \leq d_l < 0, \quad \forall l \in L,$ (13)

then $\quad (Rf)_{N(\mathcal{L})} = (Rf)_{F(\mathcal{L}_1)} = (Rf)_{F(\mathcal{L}_{1c})}.$ (13a)

**Proof.** In the case (13), the statement (13a) is obtained by a trivial generalization of statements 1–4 from [3] (pp. 119-125). Moreover, $d_l < 0$ and $d_l \geq -D$ are used to prove the analogue of statements 3 and 4 from [3]. In the process of proof, the inequalities for integer parts are used:

$$[a + b] \geq [a] + [b],$$
$$[a - b] \leq [a] - [b].$$

In case (12), it is enough in statements 1-3 of [3] to replace the word "proper" with the phrase "strongly connected" to obtain (12a).

We formulate the convergence theorem. We consider proper correlators (3) with integer ultraviolet dimensions. Below we use the notation $v = |V| - 1$, where $|v|$ is the number of vertices of the graph.

**Theorem.** *A connected graph $\Gamma$ is given that represents the Feynman integral with proper $\varepsilon$-correlators (3), whose UV dimensions (5) are integers. If we take $N(\mathcal{L})$ as the renormalization array, then*

*a) for $\varepsilon \in (0, 2\pi)$ and all $q \in \mathbf{R}^{Dv}$, the renormalized Feynman integral exists and generates*

$$R_N(\mathcal{L})(\varepsilon) = R(\varepsilon) \in \operatorname{reg} S'(\mathbf{R}^{Dv}),$$

*b) in $S'(\mathbf{R}^{Dv})$ there exist a limit*

$$\lim_{\varepsilon \to +0} R(\varepsilon), \tag{14}$$

*c) which is covariant.*

**Proof.** a) The proof of the existence of the renormalized integral fits entirely into the scheme developed in [3]: the integral is divided into sectors, classes of equivalent nests are distinguished, $R$ - operation is written using integro-differential operators with respect to the variables $\sigma_l$. The original integral is divided into the sum of $(|L|!)^2$ terms of the form

$$\mathcal{P}(q, \varepsilon) \int_\Delta du \int_0^\infty dt \left\{ \left( \prod_{l=1}^{|L|} f_l \left( ie^{-i\varepsilon/2} t \, \varphi_l(u) \right) t^a \left( \prod_{l=1}^{|L|-1} u_l^{a_l} \right) \tilde{Q}(u) \right) \right\} \times$$

$$\times \exp\left(i \, e^{-i\varepsilon/2} \, t \, \mathcal{A}(q, u, \varepsilon)\right) \times \exp\left(-ie^{-i\varepsilon/2} \, t \left( m_{|L|}^2 + \sum_{l=1}^{|L|-1} m_l^2 \varphi_l(u) \right)\right), \tag{15}$$

where $a, a_l$ are half-integers, integration over $u$ goes over the compact set $\Delta$,

$$u = (t_1, \ldots, t_{|L|-1}, \sigma_1, \ldots, \sigma_k) = (u_1, \ldots, u_{|L|+k-1}), \qquad t = t_{|L|},$$

and

$$\varphi_l(u) = u_l \ldots u_{|L|-1},$$

$$\varphi_{|L|}(u) = 1.$$

The function $\tilde{Q}(u)$ is holomorphic in all points of $\Delta$,

$$\mathcal{A}(q, u, \varepsilon) = \sum_{i,j=1}^{v} \mathcal{A}_{ij}(u)(q_i q_j)_\varepsilon,$$

where $\mathcal{A}_{ij}(u)$ are rational functions of the parameters $u$ holomorphic at all points of $\Delta$. The compact $\Delta$ is a cube: $\Delta = [0,1]^{|L|+k-1}$. $\mathcal{P}(q, \varepsilon)$ is a polynomial in the components of $q \in \mathbf{R}^{Dv}$ with coefficients that smoothly depend on $\varepsilon$.

We introduce the following notation:

$$Q(t, u, \varepsilon) \equiv \left( \prod_{l=1}^{|L|} f_l \left( i\, e^{-i\varepsilon/2} t\, \varphi_l(u) \right) \right) t^a \left( \prod_{l=1}^{|L|-1} u_l^{a_l} \right) \tilde{Q}(u). \quad (16)$$

For $\varepsilon \in [0, 2\pi], t \in (0, +\infty); u \in (0,1)^{|L|+k-1}$ the estimate is valid:

$$|Q(t, u, \varepsilon)| \le c \left( t^{-1/2 - |L|\delta} + t^N \right) \prod_{l=1}^{|L|-1} u_l^{-1/2 - l\delta}, \quad (17)$$

where $\delta > 0$ is an arbitrary positive number; $N \ge 0$ is an integer [number]. Indeed, by virtue of Definition 1, the inequality is valid:

$$|f_l(\alpha_l)| \le c_l \left( |\alpha|^{s_l - \delta} + |\alpha|^{T_l} \right),$$

where $\delta > 0$ is arbitrary, $\alpha \in \{\operatorname{Re}\alpha \ge 0\}$, $s_l \ge -1/2$ is half-integer (since the dimension: $d_l$ in (5) is integer), $T_l$ can be chosen so that $T_l \ge s_l$.

Then

$$\left| \prod_{l=1}^{|L|} f\left( i e^{-i\varepsilon/2} t \varphi_l(u) \right) \right| \le C \prod_{l=1}^{|L|} \left( \left( t\varphi_l(u) \right)^{s_l - \delta} + \left( t\varphi_l(u) \right)^{T_l} \right) \le$$

$$\le c_1 \prod_{l=1}^{|L|} (\varphi_l(u))^{s_l - \delta} \left( t^{s_l - \delta} + t^{T_l} \right) \le$$

$$\leq c_2 \left( \prod_{l=1}^{|L|-1} u_l^{\sum_{k=1}^{l} s_k - l\delta} \right) \left( t^{\sum_{l=1}^{|L|} s_l - |L|\delta} + t^{\sum_{l=1}^{|L|} T_l} \right), \tag{18}$$

where $\delta > 0$.

Combinatorial analysis in the framework of the scheme and notations of [3] gives the following estimates:

$$a_l \geq -\frac{1}{2} + \frac{1}{2} \sum_{p(j) \leq l} \left( \left( -\sum_{k \in \gamma_j^p} 2s_k \right) + \sum_{k \in \gamma_{j-1}^p} 2s_k \right) =$$

$$-\frac{1}{2} - \sum_{p(j) \leq l} s_{p(j)} = -\frac{1}{2} - \sum_{k=1}^{l} s_k. \tag{19}$$

Here $l = 1, \ldots, |L|$ and $a_{|L|} = a$. The difference between (19) and the corresponding expressions in [3] (p. 142) arises because in our case there are no combinations $(\dim P_l - 2)$ in the powers of the parameters $\sigma$ (11), as is the case in [3], but the dimensions:

$$d_l = \dim P_l - 2 - 2s_l.$$

The estimate (17) follows from (18) and (19). As $N$, we can take

$$N = \left[ \sum_{l=1}^{|L|} T_l + a + 1 \right].$$

The following estimates are also valid:

$$\left| \exp\left( ite^{-i\varepsilon/2} \mathcal{A}(q, u, \varepsilon) \right) \right| \leq 1 \tag{20}$$

and

$$\left|\exp\left(it\, e^{-i\varepsilon/2}\left(m_{|L|}^2 + \sum_{l=1}^{|L|-1} m_l^2 \varphi_l(u)\right)\right)\right| \leq \exp\left(-\left(\sin\frac{\varepsilon}{2}\right) t\, m_{|L|}^2\right) \quad (21)$$

for $\varepsilon \in [0,\, 2\pi]$ and all $q \in \mathbf{R}^{Dv}$ and $u$. Inequality (21) is trivial, (20) follows from the inequality

$$\mathrm{Re}\left(ie^{-i\varepsilon/2}\cdot k_\varepsilon^2\right) = -\left(\sin\frac{\varepsilon}{2}\right) k_E^2 \leq 0$$

and the fact that $\mathcal{A}(q, u, \varepsilon)$ is a linear combination of squares $(\sum q_i)_\varepsilon^2$, with non-negative coefficients.

The integral in (15) by virtue of inequalities (17), (20) and (21) for $\varepsilon \in [0,\, 2\pi]$ and all $q \in \mathbf{R}^{Dv}$ exists, and so there exists an initial renormalized integral (obtained from (7) by replacing $f \to Rf$, where $Rf$ is given by formula (11)). This integral is equal to the sum of terms of the form (15):

$$R(q, \varepsilon) = \sum_{i=1}^{(|L|!)^2} \mathcal{P}_i(q, \varepsilon) Z_i(q, \varepsilon). \quad (22)$$

In (22), $Z_i(q, \varepsilon)$ are defined by integrals of the form (we omit the index $i$):

$$\int_\Delta du \int_0^\infty dt \left\{\left(\prod_{l=1}^{|L|} f_l\left(ie^{-i\varepsilon/2} t\, \varphi_l(u)\right)\right) t^a \left(\prod_{l=1}^{|L|-1} u_l^{a_l}\right) \tilde{Q}(u)\right\} \times$$

$$\times \exp\left(ie^{-i\varepsilon/2} t\mathcal{A}(q, u, \varepsilon)\right) \exp\left(-ie^{-i\varepsilon/2} t\left(m_{|L|}^2 + \sum_{l=1}^{|L|-1} m_l^2 \varphi_l(u)\right)\right) =$$

$$= Z(q, \varepsilon) \quad (15a)$$

with previously introduced notations.

By virtue of inequalities (17), (20) and (21) for $\varepsilon \in (0,\, 2\pi)$ $Z_i(q, \varepsilon)$ are smooth over $q$ in $\mathbf{R}^{Dv}$ and bounded. Then $R(q, \varepsilon)$ is "slowly integrable" (see [1]) for $\varepsilon \in (0,\, 2\pi)$ and generates

$$R(\varepsilon) = \sum_{i=1}^{(|L|!)^2} \mathcal{P}_i(.,\varepsilon) Z_i(\varepsilon) \in \mathrm{reg}\, S'(\mathbf{R}^{Dv})$$

where $Z_i(\varepsilon) \in \text{reg } S'(\mathbf{R}^{Dv})$ are generated by $Z_i(q, \varepsilon)$; $\mathcal{P}_i(., \varepsilon)$ are operators (in $S'(\mathbf{R}^{Dv})$) of multiplication by polynomials $\mathcal{P}_i(q, \varepsilon)$.

b) to prove the existence of a limit (14), it is sufficient to prove the existence of the limits in $S'(\mathbf{R}^{Dv})$):

$$\lim_{\varepsilon \to +0} Z_i(\varepsilon) = Z_i, \qquad (23)$$

then the limit (14) exists and is equal to

$$\lim_{\varepsilon \to +0} R(\varepsilon) = \sum_{i=1}^{(|L|!)^2} \mathcal{P}_i(., 0) Z_i, \qquad (24)$$

where $\mathcal{P}_i(., 0)$ are continuous operators in $S'(\mathbf{R}^{Dv})$ - operators of multiplication by $\mathcal{P}_i(q, 0)$. $\mathcal{P}_i(q, 0)$ are covariant.

So, it is enough for us to prove the existence of the limit $\varepsilon \to +0$ in $S'(\mathbf{R}^{Dv})$ for a regular generalized function given by the integral (15a).

The proof of this fact is carried out in complete analogy with the proof presented in [3] (p. 145-148).

We divide (15a) into the sum of two terms:

$$Z(q, \varepsilon) = Z(q, \varepsilon|1 - \Psi) + Z(q, \varepsilon|\Psi), \qquad (25)$$

where

$$Z(q, \varepsilon|\Psi) \equiv \int_\Delta du \int_0^\infty dt Q(t, u, \varepsilon) \Psi(\mathcal{A}(q, u, 0)) \times$$

$$\times \exp\left(ie^{-i\varepsilon/2} t(\mathcal{A}(q, u, \varepsilon) - m^2 - M^2(u))\right)$$

and a similar expression for $Z(q, \varepsilon|1 - \Psi)$, obtained by replacing $\Psi \to (1 - \Psi)$.

Here $\Psi(x)$ is a smooth function on $\mathbf{R}$ equal to zero for $x < m^2/3$ and unity for $x > 2m^2/3$; $m^2 = m_{|L|}^2$ и $M^2(u) = \sum_{l=1}^{|L|-1} m_l^2 \varphi_l(u)$. Both functions in (25) generate regular generalized functions for $\varepsilon \in (0, 2\pi)$

$$Z(\varepsilon|\Psi), Z(\varepsilon|1 - \Psi) \in \text{reg } S'(\mathbf{R}^{Dv}).$$

We give one useful relation used in the following:

$$ie^{-i\varepsilon/2}(\mathcal{A}(q,u,\varepsilon) - \tilde{m}^2) = i\left(\cos\frac{\varepsilon}{2}\right)(\mathcal{A}(q,u,0) - \tilde{m}^2)$$

$$-\left(\sin\frac{\varepsilon}{2}\right)(\mathcal{A}_E(q,u) + \tilde{m}^2), \quad (26)$$

where $\mathcal{A}_E(q,u) = -\mathcal{A}(q,u,\pi)$ is the sum of Euclidean squares with non-negative coefficients.

Consider $Z(q,\varepsilon|1-\Psi)$. The presence of the factor $\left(1 - \Psi(\mathcal{A}(q,u,0))\right)$ makes it possible for $\varepsilon \in (0, \pi)$ to rotate the integration contour $t \to -it$. Indeed, $q$ with $\mathcal{A}(q,u,0) < 2m^2/3$ are essential, then for $\varepsilon \in (0, 2\pi)$, by virtue of (26), we obtain:

$$\mathrm{Re}\left(ie^{-i\varepsilon/2}(\mathcal{A}_\varepsilon - m^2 - M^2(u))\right) = -\left(\sin\frac{\varepsilon}{2}\right)(\mathcal{A}_E + m^2 + M^2(u)) < 0; \quad (27)$$

$$\mathrm{Im}\left(ie^{-i\varepsilon/2}(\mathcal{A}_E - m^2 - M^2(u))\right) = \left(\cos\frac{\varepsilon}{2}\right)(\mathcal{A}_M - m^2 - M^2(u)) < 0,$$

where $\mathcal{A}_M = \mathcal{A}(q,u,0)$, $\mathcal{A}_E = \mathcal{A}_E(q,u)$, $\mathcal{A}_\varepsilon = \mathcal{A}(q,u,\varepsilon)$. By virtue of (27), rotation of the integration contour with respect to $t$ is possible by an angle $-\pi/2 \leq \varphi \leq 0$.

Then

$$Z(q,\varepsilon|1-\Psi) = (-i)\int_\Delta du \int_0^\infty dt\, Q(-it, u, \varepsilon) \times$$

$$\times \Psi(\mathcal{A}(q,u,0)) \times \exp\left(e^{-i\varepsilon/2}t(\mathcal{A}(q,u,\varepsilon) - m^2 - M^2(u))\right), \quad (28)$$

where $\varepsilon \in (0, \pi)$.

The integral in (28) also exists for $\varepsilon = 0$ and defines a bounded smooth function $Z(q|1-\Psi)$, generating

$$Z(1-\Psi) \in \mathrm{reg}\, S'(\mathbf{R}^{D\nu}).$$

Moreover, in $S'(\mathbf{R}^{D\nu})$

$$\lim_{\varepsilon\to+0} Z(\varepsilon|1-\Psi) = Z(1-\Psi). \quad (29)$$

Relation (29) (as well as similar relations below) are easily proved by applying the Lebesgue theorem [2] (p. 302) to the mixed integral $\int dq du dt\{...\}$, obtained from

$$\langle Z(\varepsilon|1-\Psi), \varphi \rangle = \int dq Z(q, \varepsilon|1-\Psi)\varphi(q),$$

where $\varphi \in S(\mathbf{R}^{Dv})$. (The Lebesgue theorem formulated in [2] for sequences is also valid for $\varepsilon$-networks).

Consider the second term in (25): $Z(q, \varepsilon|\Psi)$. We divide it into the sum of two terms:

$$Z(q, \varepsilon|\Psi) = Z_1(q, \varepsilon|\Psi) + Z_2(q, \varepsilon|\Psi), \qquad (30)$$

where

$$Z_1(q, \varepsilon|\Psi) = \int_\Delta du \int_0^1 dt \, Q(t, u, \varepsilon) \, \Psi(\mathcal{A}(q, u, 0)) \times$$

$$\times \exp\left(ie^{-i\varepsilon/2} t(\mathcal{A}(q, u, \varepsilon) - m^2 - M^2(u))\right), \qquad (31)$$

and similarly

$$Z_2(q, \varepsilon|\Psi) = \int_\Delta du \int_1^\infty dt \, Q(t, u, \varepsilon) \, \Psi(\mathcal{A}(q, u, 0)) \times$$

$$\times \exp\left(ie^{-i\varepsilon/2} t(\mathcal{A}(q, u, \varepsilon) - m^2 - M^2(u))\right). \qquad (32)$$

For $\varepsilon \in [0, 2\pi]$ relation (31) defines an $\varepsilon$-network of regular generalized functions from $S'(\mathbf{R}^{Dv})$ $Z_1(\varepsilon|\Psi)$, which is continuous at the point $\varepsilon = 0$:

$$\lim_{\varepsilon \to +0} Z_1(\varepsilon|\Psi) = Z_1(0|\Psi). \qquad (33)$$

Consider $Z_2(\varepsilon|\Psi)$ - a network of regular generalized functions from $S'(\mathbf{R}^{Dv})$, generated by (32) ($\varepsilon \in (0, 2\pi)$). The following representation is valid:

$$Z_2(\varepsilon|\Psi) = \hat{Q} \cdot X(\varepsilon|\Psi) - Y(\varepsilon|\Psi), \qquad (34)$$

where $X(\varepsilon|\Psi) \in \text{reg } S'(\mathbf{R}^{Dv})$ is generated by a smooth bounded function

$$X(q, \varepsilon|\Psi) = \int_\Delta du \int_1^\infty dt Q(t, u, \varepsilon) t^{-N-2} \Psi(\mathcal{A}(q, u, 0)) \times$$

$$\times \left(ie^{-i\varepsilon/2}\mathcal{A}(q,u,\varepsilon)\right)^{-N-2}\exp\left(ie^{-i\varepsilon/2}t(\mathcal{A}(q,u,\varepsilon) - m^2 - M^2(u))\right), \tag{35}$$

$\widehat{Q}$ is continuous operator in $S'(\mathbf{R}^{Dv})$:

$$\widehat{Q} = \left(\frac{1}{2}q\frac{\partial}{\partial q} + 1\right)\ldots\left(\frac{1}{2}q\frac{\partial}{\partial q} + N + 2\right), \tag{36}$$

here

$$q\frac{\partial}{\partial q} = \sum_{i=1}^{v}\sum_{\alpha=0}^{D-1} q_{i\alpha}\frac{\partial}{\partial q_{i\alpha}}.$$

$Y(\varepsilon|\Psi) \in \operatorname{reg} S'(\mathbf{R}^{Dv})$ порождено $Y(q,\varepsilon|\Psi)$ — суммой интегралов, получаемых при действии дифференциального оператора $\widehat{Q}$ (36) на $X(q,\varepsilon|\Psi)$ — тех и только тех слагаемых, в которых хотя бы одна производная из (36) действует на $\Psi(\mathcal{A}(q,u,0))$. Указанное разбиение корректно. Действительно, в силу наличия в (35) множителя $\Psi(\mathcal{A}(q,u,0))$ существенны те $q$, для которых

$Y(\varepsilon|\Psi) \in \operatorname{reg} S'(\mathbf{R}^{Dv})$ is generated by $Y(q,\varepsilon|\Psi)$ – the sum of integrals obtained under the action of the differential operator $\widehat{Q}$ (36) on $X(q,\varepsilon|\Psi)$ - those and only those terms in which at least one derivative from (36) acts on $\Psi(\mathcal{A}(q,u,0))$. The specified splitting is correct. Indeed, due to the presence of the factor $\Psi(\mathcal{A}(q,u,0))$ in (35), those $q$ are essential for which

$$\mathcal{A} = \mathcal{A}(q,u,0) \geq m^2/3,$$

then

$$\mathcal{A}_E = \mathcal{A}_E(q,u) \geq \mathcal{A}(q,u,0) \geq m^2/3.$$

By virtue of (26)

$$\left|i\,e^{-i\varepsilon/2}\mathcal{A}(q,u,\varepsilon)\right| = \left(\mathcal{A}_M^2\cos^2\frac{\varepsilon}{2} + \mathcal{A}_E^2\sin^2\frac{\varepsilon}{2}\right)^{1/2} \geq m^2/3. \tag{37}$$

Then, due to of inequalities (17), (20), (21) and (37), integral (35) exists for $\varepsilon \in [0,\ 2\pi]$ and all $q$ and generates an ε-network ($\varepsilon \in [0,\ 2\pi]$)

$$X(\varepsilon|\Psi) \in \operatorname{reg} S'(\mathbf{R}^{Dv})$$

continuous at the point $\varepsilon = 0$:

$$\lim_{\varepsilon \to +0} X(\varepsilon|\Psi) = X(0|\Psi). \tag{38}$$

Representation (34) is easily verified. As for $Y(q, \varepsilon|\Psi)$, here we have the sum of the integrals in which the derivatives of $\Psi$ appear. The derivatives of $\Psi$ have supports contained in $[m^2/3, 2m^2/3]$. As in the case of $Z(q, \varepsilon|1 - \Psi)$, for $\varepsilon \in (0, \pi)$ we can rotate the integration contour over $t$: $t \to -it$ and prove the existence of the limit

$$\lim_{\varepsilon \to +0} Y(\varepsilon|\Psi) = Y(\Psi). \tag{39}$$

By virtue of (25), (29), (30), (33), (34), (38), (39) and the continuity of the operator $\hat{Q}$, there exists a limit

$$\lim_{\varepsilon \to +0} Z(\varepsilon) = Z(1 - \Psi) + Z_1(0|\Psi) + \hat{Q}X(0|\Psi) - Y(\Psi) = Z. \tag{40}$$

That is, the existence of limits (71) and (72) is proved.

c) Explicit expressions for $Z(1 - \Psi)$, $Z_1(0|\Psi)$, $X(0|\Psi)$ and $Y(\Psi)$ are covariant; therefore, $Z$ in (40) is covariant. Then (24) is covariant due to the covariance of $Z_i$ and $\mathcal{P}_i(q, 0)$.

**4. Possible applications in gravity.** So far, we have considered the flat case. However, the regularization of the ε-metric can be generalized to the case of curved space-time. Indeed, let a pseudo-Euclidean metric be given

$$g = g_{\mu\nu}(x) dx^\mu \otimes dx^\nu \tag{41}$$

and D-bein is given

$$e^a = e^a_\mu(x) dx^\mu, \tag{42}$$

locally leading (41) to the canonical form

$$g = \eta_{ab} e^a \otimes e^b \tag{43}$$

and in coordinates

$$g_{\mu\nu}(x) = \eta_{ab} e^a_\mu(x) e^b_\nu(x).$$

Consider the ε-metric

$$g^\varepsilon = \eta^\varepsilon_{\alpha\beta} e^\alpha \otimes e^\beta \tag{44}$$

and accordingly in the coordinates

$$g^\varepsilon_{\mu\nu}(x) = \eta^\varepsilon_{\alpha\beta} e^\alpha_\mu(x) e^\beta_\nu(x),$$

where $\eta^\varepsilon_{\alpha\beta}$ is defined in (1).

Such an ε-metric is defined not by $g$ (41), but by D-bein (42), leading (41) to its normal (canonical) form. For $\varepsilon = \pi$, we obtain

$$g^E = -g^{\varepsilon=\pi} = \sum_{\alpha=0}^{D-1} e^\alpha \otimes e^\alpha$$

- the Euclidean metric accompanying (43).

Using the ε-metric (44), one can correctly justify the Schwinger-DeWitt proper time formalism [4]. So, for example, for the inverse operator

$$(-i0 + m^2 + \Box_0)^{-1}, \quad m^2 > 0,$$

where (for $D = 4$)

$$\Box_0(.) = \frac{1}{\sqrt{-g(x)}} \partial_\mu \left(\sqrt{-g(x)} g^{\mu\nu}(x) \partial_\nu(.)\right).$$

Proper time representation looks like

$$(-i0 + m^2 + \Box_0)^{-1} = \lim_{\varepsilon \to +0} i e^{-i\varepsilon/2} \int_0^\infty ds \, \exp\left(-s \, i \, e^{-i\varepsilon/2} (\Box_0^\varepsilon + m^2)\right).$$

Here

$$\Box_0^\varepsilon(.) = \frac{1}{\sqrt{-g_\varepsilon(x)}} \partial_\mu \left(\sqrt{-g_\varepsilon(x)} g^{\mu\nu}_\varepsilon(x) \partial_\nu(.)\right) = \frac{1}{e(x)} \partial_\mu \left(e(x) g^{\mu\nu}_\varepsilon(x) \partial_\nu(.)\right),$$

where

$$\left(g^{\mu\nu}_\varepsilon(x)\right) = \left(g^\varepsilon_{\mu\nu}(x)\right)^{-1} = \left(\eta^{\alpha\beta}_\varepsilon e^\mu_\alpha(x) e^\nu_\beta(x)\right)$$

and

$$g_\varepsilon(x) = \det\left(g^{\mu\nu}_\varepsilon(x)\right) = -e^{-i\varepsilon} e^2(x),$$

$$e(x) = \det\left(e^\alpha_\mu(x)\right),$$

and $\sqrt{...}$ is the branch of the square root with the cut $[0, +\infty)$, normalized by the condition: $\sqrt{1+i0} = -1$.

Using the ε-metric (44), the transition to the Euclidean theory is correctly carried out. Perhaps the use of (44) will prove to be useful in connection with Hawking's hypothesis about the Euclidean nature of the initial singularity and tunneling into the pseudo-Euclidean world [5]. A number of physical questions remain open: should we consider different signature sectors of a non-degenerate metric when quantizing, or should we consider a pseudo-Euclidean (or Euclidean in Hawking's formalism) sector? Are signature phase transitions possible? Are transitions possible with a change in the arrow of time? Do domains with a Euclidean or 2 + 2 signature exist in the Universe? etc.

Perhaps the introduction of a complex metric will be useful not only from a methodological point of view, but also from a physical point of view.

**5. Conclusion.** The regularization of pseudo-Euclidean singularities introduced in [1] using the complex ε-metric has several advantages compared to covariant regularization [3]. It preserves some properties of the Euclidean theory for $\varepsilon \in (0, 2\pi)$: at large momenta, ε-correlators usually behave like Euclidean correlators, and so the convergence of Euclidean integrals implies the convergence of the integrals in the ε-metric; the regularity of ε-correlators in the coordinate representation, as a rule, is guaranteed by regularity [of correlators] in the Euclidean case (see [1]). In addition, if the integral "good in indices" in the covariant regularization, generally speaking, exists only in the $\alpha$-representation (see [1]), then in the ε-metric this integral also exists in the momentum representation. Thus, using the ε-metric, the transition to the $\alpha$-representation is justified. The only drawback of the proposed regularization - non-covariance - is of an intermediate nature: in the limit $\varepsilon = +0$, covariant generalized functions are obtained[**].

All of the above gives reason to believe that regularization by a complex metric is an integral part of constructing a perturbation theory for a quantum field theory in the space of a pseudo-Euclidean (or any other alternating) signature.

___

[**]   At the moment some phrases of this paragraph sound as little bit naive. (Year 2020.)

LITERATURE

1. Ivashchuk V.D., [Regularization by ε-metric. I] , Izv. AN MSSR, Ser. fiz.-tekhn. i mat. nauk, 1987, № 3, p. 8-17 [in Russian]; arXiv: 1902.03152.

2. Kolmogorov A. N., Fomin S. V. Elements of the theory of functions and functional analysis. M.: Nauka, 1981. [in Russian]

3. Zavyalov O. I. Renormalized Feynman diagrams. M.: Nauka, 1979. [in Russian]

4. Schwinger J. Phys. Rev. 1951. **82**. P. 664.
DeWitt B. S. Dynamical Theory of Groups and Field. Gordon and Breach, N.-Y., 1965.

5. Hawking S. Cambridge University preprint. 1982. [***)]

---

[***)] Apparently, the author here had in mind the preprint version of the famous article:

Hartle, J.; Hawking, S. (1983). "Wave function of the Universe". Physical Review D **28** (12): 2960.

Additional references (added in 2020)

**a. BPHZ-theorem**

[1a.] N. N. Bogoliubow und O. S. Parasiuk, "Über die Multiplikation der Kausalfunktionen in der Quantentheorie der Felder", Acta Mathematica, B. 97, S. 227–266 (1957); doi:10.1007/BF02392399.

[2a.] K. Hepp, "Proof of the Bogoliubov-Parasiuk Theorem on Renormalization", Communications in Mathematical Physics, V. 2, N. 1, P. 301–326 (1966); doi:10.1007/BF01773358.

[3a.] W. Zimmermann, "Convergence of Bogoliubov's Method of Renormalization in Momentum Space", Communications in Mathematical Physics, V. 15, No. 3, P. 208–234 (1969); doi:10.1007/BF01645676.

**b. Some related papers on Wick rotation**

[1b] P. Candelas and D.J. Raine, "Feynman propagator in curved space-time", Phys. Rev. D 15 (1977) 1494; doi:10.1103/PhysRevD.15.1494.

[2b] G.W. Gibbons, "The Einstein action of Riemannian metrics and its relation to quantum gravity and thermodynamics", Phys. Lett. A 61 (1977) 3; doi:10.1016/0375-9601(77)90244-4

[3b] V.D. Ivashchuk, "(1/N) -theory of Perturbations in the Space with Regularized Minkowski Metric" (1982) [in Russian] (unpublished).

[4b] M. Visser, "How to Wick rotate generic curved spacetime" (1991), Gravity Research Foundation essay; arXiv:1702.05572.

[5b] J. Greensite, "Stability and signature in quantum gravity".

[6b] J. Greensite, "Dynamical origin of the Lorentzian signature of space-time", Phys. Lett. B 300 (1993) 34; gr-qc/9210008.

[7b] A. Carlini and J. Greensite, "Why is space-time Lorentzian?", Phys. Rev. D 49 (1994) 866; gr-qc/9308012.

[8b] J. Greensite, "Quantum mechanics of space-time signature", Acta Phys. Polon. B 25 (1994) 5.

[9b] E. Elizalde, S.D. Odintsov and A. Romeo, **"Dynamical determination of the metric signature in spacetime of non-trivial topology"**, Class. Quant. Grav., 11, No. 4 (1994) L61-L68; hep-th/9312132.

[10b] A. Carlini and J. Greensite, "Square root actions, metric signature, and the path integral of quantum gravity", Phys. Rev. D 52 (1995) 6947; gr-qc/9502023.

[11b] S.A. Hayward, "Complex lapse, complex action and path integrals", Phys. Rev. D 53 (1996) 5664; gr-qc/9511007.

[12b] V.D. Ivashchuk, "Wick rotation, regularization of propagators by a complex metric and multidimensional cosmology", Grav. Cosmol. 3 (1997) 8-16; arXiv:gr-qc/9705008.

[13b] J.F. Barbero G., "From Euclidean to Lorentzian general relativity: The real way", Phys. Rev. D 54 (1996) 1492; gr-qc/9605066.

[14b] J. Ambjorn, J. Jurkiewicz and R. Loll, "Quantum gravity, or the art of building spacetime", In *Oriti, D. (ed.): Approaches to quantum gravity* 341-359; hep-th/0604212.

[16b] B.P. Kosyakov, "Black holes: Interfacing the classical and the quantum", Found. Phys. 38 (2008) 678; arXiv:0707.2749 [gr-qc].

[17b] F. Girelli, S. Liberati and L. Sindoni, "Emergence of Lorentzian signature and scalar gravity", Phys. Rev. D 79 (2009) 044019; arXiv: 0806.4239 [gr-qc].

[18b] A. White, S. Weinfurtner and M. Visser, "Signature change events: A Challenge for quantum gravity?", Class. Quant. Grav. 27 (2010) 045007; arXiv: 0812.3744 [gr-qc].

[19b] C. Helleland and S. Hervik, "A Wick-rotatable metric is purely electric", arXiv:1504.01244 [math-ph].

[20b] J. Samuel, "Wick Rotation in the Tangent Space", Class. Quant. Grav. 33 (2016) 015006; arXiv:1510.07365 [gr-qc].

[21b] F. Gray, "Black hole radiation, greybody factors, and generalised Wick rotation", MSc thesis, 2016, Victoria University of Wellington. http://hdl.handle.net/10063/5148

[22b] C. Helleland and S. Hervik, "Wick rotations and real GIT" ; arXiv: 1703.04576.

[23b] A. Baldazzi, R. Percacci and V. Skrinjar, "Wicked metrics", Class. Quant. Grav. 36 (2019) 105008; arXiv:1811.03369 .